\documentstyle[aps,prb,preprint]{revtex}

\include{psfig}

\begin{document}
\draft
\title{Theory of the field-effect mobility in 
amorphous organic transistors}
\author{M. C. J. M. Vissenberg$^{1,2,}$\cite{email} 
and M. Matters$^{1,}$\cite{email}}
\address{$^1$Philips Research Laboratories, 5656 AA Eindhoven, 
The Netherlands\\
$^2$Instituut--Lorentz, University of Leiden, 2300 RA Leiden, The Netherlands }
\date{\today}
\maketitle

\begin{abstract}
The field-effect mobility in an organic thin-film transistor is studied 
 theoretically. 
From a percolation model  of hopping between localized states and 
a transistor model an analytic expression for the field-effect 
mobility is obtained. 
The theory is applied to describe the experiments 
by Brown et al. [Synth.\ Met.\ {\bf 88}, 37 (1997)] on 
solution-processed amorphous organic transistors, 
made from a polymer (polythienylene vinylene) and 
from a small molecule (pentacene). 
Good agreement is obtained, both with respect to the 
gate-voltage and the temperature dependence of the mobility.
\end{abstract}

\pacs{PACS numbers: 72.80.Le, 85.30.Tv\\[1ex]
Submitted to Phys. Rev. B, 13 jan. 1998}

Over the last decade the use of organic $p$-type semiconductors in
field-effect transistors has gained considerable interest due to 
their potential application in low-cost integrated circuits. 
Most effort has been put into increasing 
the hole mobility of the semiconductor 
and increasing the on-off ratio of the field-effect 
transistor by optimising existing materials and by applying new materials. 
Mobilities as high as 0.7 cm$^2$V$^{-1}$s$^{-1}$ and 
on-off ratios of 10$^8$ have recently been reported in thin-film
transistors of evaporated pentacene.~\cite{gundlach} 
Furthermore, attention has been focused on the improvement of
 the processability of these materials by using directly 
soluble~\cite{bao} or precursor organic semiconductors.~\cite{overview} 
Besides the technical applicability of organic semiconductors, 
their electronic and structural
properties have been the subject of investigation as well. 
Interesting questions like the connection 
between molecular order and hole mobility in conjugated oligomers and polymers
have been addressed.~\cite{gundlach,garnier,laquindanum,dodabalapur} 

Experiments have indicated that the field-effect mobility of holes 
in organic transistors depends on the temperature as well as on the 
applied gate bias.~\cite{overview,horowitz} This has been described by
Horowitz {\it et al.} using a 
multiple trapping and release model.~\cite{horowitz}
In this model the assumption is made that most of the charge carriers 
are trapped in localized states. 
Then the amount of (temporarily) released charge carriers to an extended-state 
transport level (the valence band for classical 
{\it p}-type semiconductors) depends on the energy level of the 
localized states, the temperature, and the gate voltage. 
However, while extended-state transport may occur in highly ordered 
vacuum-evaporated molecular films,~\cite{horowitz} we do not expect it  
to play a role in amorphous organic films,~\cite{overview} 
where the charge carriers are strongly localized.

In the present paper, we derive a theory for the field-effect mobility 
in amorphous organic transistors, where the 
charge transport is governed by hopping, 
i.e. the thermally activated tunneling 
of carriers between localized states, 
rather than by the activation of carriers to a transport level. 
We use the concept of variable range hopping (VRH), i.e.\ a carrier 
may either hop over a small distance  with a high activation energy 
or hop over a long distance with a low activation energy. 
The temperature dependence of the carrier transport in 
such a system is strongly dependent on the density of localized states. 
In a field-effect transistor, an applied gate voltage 
gives rise to the accumulation of charge in the region of the 
semiconducting layer that is close to the insulator. 
As these accumulated charge carriers fill the lower-lying states of the 
organic semiconductor, any additional charges in the accumulation layer 
will occupy states at relatively high energies.  
Consequently, these additional charges will---on average---require less 
activation energy to hop away to a neighboring site. 
This results in a higher mobility with increasing gate voltage. 

This paper is organized as follows. First, we study the influence of 
temperature and the influence of the filling of states 
on the conductivity of a VRH system with an exponential distribution 
of localized-state energies. 
Using percolation theory, we find an analytic expression for 
the conductivity.
This expression is then used to derive the field-effect mobility of the 
carriers when the material is applied in a transistor.  
Finally, our result is used to interpret the experimentally observed 
temperature and gate-voltage dependence of the field-effect mobility 
in both a pentacene and a 
polythienylene vinylene (PTV) organic thin-film transistor.~\cite{overview} 

Let us first derive an expression for the conductivity as a function of 
temperature $T$ and charge carrier density.
At low carrier densities and low $T$, 
the transport properties are determined 
by the tail of the density of (localized) states (DOS). 
Our model is based on the following exponential DOS, 
\begin{equation}
g(\epsilon) = \frac{N_{\rm t}}{ k_{\rm B} T_0 } 
\exp\left( \frac{\epsilon}{ k_{\rm B } T_0} \right) ,  \:
  (-\infty < \epsilon \le 0) 
\label{eq:DOS}
\end{equation}
where $N_{\rm t}$ is the number of states per unit volume, 
$k_{\rm B}$ is Boltzmann's constant, and $T_0$ is a parameter that indicates  
 the width of the exponential distribution. 
We take $g(\epsilon) =0$ for positive values of $\epsilon$. 
We do not expect the results to be qualitatively different 
for a different choice of $g(\epsilon)$, as long as $g(\epsilon)$ 
increases strongly with $\epsilon$. 

Let the system be filled with charge carriers, such that a fraction 
$\delta \in [0,1]$ of the localized states is occupied by a carrier, 
i.e.\ such that the 
density of charge carriers is $\delta N_{\rm t}$.
In equilibrium, the energy distribution of the carriers is given by 
the Fermi-Dirac distribution $f (\epsilon, \epsilon_{\rm F})$, where 
$\epsilon_{\rm F}$ is the Fermi energy (or chemical potential). 
For a given carrier occupation $\delta$, the position of the Fermi energy 
$\epsilon_{\rm F}$ is fixed by the condition
\begin{eqnarray}
\delta & = & \frac{1}{N_{\rm t}} 
\int d \epsilon \: g(\epsilon) f (\epsilon, \epsilon_{\rm F}) 
\nonumber \\
 & \simeq & \exp{\left( \frac{\epsilon_{\rm F}}{k_B T_0} \right)} 
\Gamma( 1-T/T_0) \Gamma (1+T/T_0) , 
\label{chempot}
\end{eqnarray}
where $\Gamma(z) \equiv \int_0^{\infty} dy \exp{(-y)} y^{z-1}$. 
In Eq.~(\ref{chempot}), we have used the 
assumption that most carriers occupy the sites with energies $\epsilon \ll 0$, 
i.e., $-\epsilon_{\rm F} \gg k_{\rm B} T_0$. 
This condition is fulfilled when $\delta$ and $T$ are low. 
Note that at $T=0$ the gamma functions are unity and the carrier 
density is given by the density of states with energies lower than 
$\epsilon_{\rm F}$.
Our approximate expression~(\ref{chempot}) breaks down  
at temperatures $T \gtrsim T_0$, where $\Gamma( 1-T/T_0)$ diverges. 
At such temperatures our assumption that only the tail 
of the DOS is important no longer holds,  
as the majority of the carriers is located close to $\epsilon =0$. 

The transport of carriers is governed by the hopping 
of carriers between localized states, which is strongly 
dependent on the hopping distances as well as the energy 
distribution of the states. 
At low bias, the system can be described as a 
resistor network.~\cite{kirkpatrick,ambegaokar} 
In this case, one can assign a conductance 
$G_{ij} = G_0 e^{- s_{ij}}$ between site $i$ and site $j$ 
where 
\begin{equation} 
s_{ij} = 2 \alpha R_{ij} + \frac{\left| \epsilon_i - \epsilon_{\rm F} \right| 
+ \left| \epsilon_j - \epsilon_{\rm F} \right| + 
\left| \epsilon_i - \epsilon_j \right|}{ 2 k_B T} .
\label{eq:macht}
\end{equation}
Here, the first term on the right-hand-side 
describes the tunneling process, 
which depends on the overlap of electronic wavefunctions of the 
sites $i$ and $j$. In a lowest-order approximation, 
this tunneling process may be characterized by 
the distance $R_{ij}$ between the sites and an effective
overlap parameter $\alpha$. The second term in Eq.~(\ref{eq:macht}) 
takes into account the activation energy for a hop upwards in energy and 
the occupational probabilities of the sites $i$ and $j$. 

According to percolation theory,~\cite{ambegaokar,sahimi} 
the conductivity of the system is given by 
\begin{equation}
 \sigma=\sigma_0 e^{-s_c}, 
\label{eq:critcond}
\end{equation}
where $\sigma_0$ is an (unknown) prefactor and $s_c$ is the exponent of 
the critical percolation conductance $G_c=G_0 e^{-s_c}$. 
This $G_c$ is determined as follows. 
Take a reference conductance $G$ 
and remove all conductances with $G_{ij}<G$. 
For high $G$, 
the remaining conductances form isolated clusters. 
These clusters increase in size with decreasing $G$.
The critical percolation conductance 
$G_c$ is defined as the value of $G$ 
at which the first infinite cluster is formed. 
This $G_c$ determines the conductivity~(\ref{eq:critcond}), 
since it is the most difficult step 
required for transport through a macroscopic system. 
The onset of percolation, i.e., the occurrence of the 
first infinite cluster, is usually determined 
by calculating the average number of bonds (conductances with $G_{ij}>G$) 
per site in the largest cluster. 
This number of bonds $B$ increases with decreasing $G$, until, at the 
onset of percolation, a critical number $B_c$ is reached. 
For a three-dimensional amorphous system, this number is given by
$B_{\rm c}\simeq 2.8$.~\cite{sahimi,pike}
The percolation criterion $B(G=G_c)=B_c$ can be  written as
\begin{equation}
B_c = \frac{N_{\rm b}}{N_{\rm s}} , 
\label{eq:perc1}
\end{equation}
where $N_{\rm b}$ is the density of bonds  
and $N_{\rm s}$ is the density of sites in the percolating system.
The density of bonds is given by 
\begin{equation}
N_{\rm b} = \int d{\bf R}_{ij} d \epsilon_i d \epsilon_j g(\epsilon_i)
g(\epsilon_j) \theta \left( s_c -s_{ij} \right) . 
\label{eq:perc2}
\end{equation}
The density of sites $N_{\rm s}$ can be estimated by 
excluding all sites that cannot possibly belong to an 
infinite cluster~\cite{butcher}
\begin{equation}
N_{\rm s} = \int d \epsilon g(\epsilon)
\theta \left(s_c k_B T - \left| \epsilon - \epsilon_{\rm F} \right| \right) .
\label{eq:perc3}
\end{equation}
Substituting Eqs.~(\ref{eq:DOS}) and~(\ref{eq:macht}) into 
Eqs.~(\ref{eq:perc1}), (\ref{eq:perc2}), and~(\ref{eq:perc3}), we obtain 
the percolation criterion for our system, 
\begin{equation}
B_c \approx \pi {\left( \frac{T_0}{2 \alpha T} \right)}^3
 N_{\rm t} \exp{\left( \frac{\epsilon_{\rm F} + s_c k_B T}{k_B T_0} \right)} , 
\label{eq:bonds}
\end{equation}
where we have assumed that the site positions are random, 
that most of the hopping takes place between tail states 
($-\epsilon_{\rm F} \gg k_B T_0$), 
and that the maximum energy hop is large ($s_c k_B T \gg k_B T_0$). 
We note that our percolation criterion~(\ref{eq:bonds}) 
is, up to a numerical factor, in agreement with 
previous results~\cite{grunewald,shapiro}  
where somewhat different approaches have been used to describe VRH in 
an exponential band tail.

Combining Eqs.~(\ref{chempot}),~(\ref{eq:critcond}) and~(\ref{eq:bonds}),
the expression for the conductivity as a function of the 
occupation $\delta$ and the temperature $T$ is obtained, 
\begin{equation}
\sigma(\delta,T)  =  \sigma_0 
\left(\frac{\pi N_{\rm t} \: \delta \: \left(T_0/T\right)^3}
{\left(2 \alpha\right)^3 B_{\rm c} \:\Gamma(1-T/T_0) \:\Gamma(1+T/T_0)}
\right)^{T_0/T}.  
\label{eq:sigmaqT}
\end{equation}
Note that the conductivity has an Arrhenius-like temperature dependence 
$\sigma \sim \exp{\left[ - E_{\rm a} / ( k_{\rm B} T ) \right]}$, 
with an activation energy $E_{\rm a}$ that is weakly (logarithmically) 
temperature dependent. 
This is in strong contrast with the well-known Mott 
formula for VRH in a constant DOS, where 
$\sigma \sim \exp{\left[ - {(T_1/T)}^{1/4}  \right]}$.~\cite{mott} 
The temperature dependence of the Mott formula is a consequence 
of hopping over far distances and hopping to high energies 
being equally important. 
In an exponential DOS, however, the characteristic hop is an activated jump, 
since there are much more available states at higher energies. 
In fact, it has been shown that hopping in an exponential DOS 
can be effectively described in terms of activation from the Fermi 
energy to a specific transport energy.~\cite{monroe,baranovskii}
This explains the Arrhenius behavior of Eq.~(\ref{eq:sigmaqT}).

In our expression~(\ref{eq:sigmaqT}), the conductivity 
increases superlinearly with the density of carriers 
($\sigma \sim {\delta}^{T_0/T}$).  
This is due to the filling of localized states: 
an increase in the carrier density gives rise to an 
increase in the average energy, thus facilitating an activated 
jump to the transport energy mentioned above. 
When the filling of states is not taken into account, i.e. 
when Boltzmann statistics instead of Fermi-Dirac 
statistics is used in Eq.~(\ref{chempot}), 
the conductivity is simply proportional to the 
density of carriers.~\cite{grunewald,shapiro} 

We now apply the obtained conductivity~(\ref{eq:sigmaqT}) 
to describe the field-effect mobility $\mu_{\rm FE}$ in a transistor.
In bulk material, the  mobility $\mu$ of the 
charge carriers is given by $\mu = \sigma(\delta,T) / (e \delta  N_{\rm t})$. 
In a transistor, however, the charge density  
is not uniform, but it decreases with the distance $x$ 
from the semiconductor-insulator interface. 
According to Eq.~(\ref{chempot}), the occupation $\delta(x)$ 
depends on the distance $x$ through the gate-induced potential $V(x)$,
\begin{equation}
\delta(x) = \delta_0 \exp\left(\frac{e V(x)}{k_{\rm B} T_0}\right),
\label{eq:qvsx}
\end{equation}
where $\delta_0$ is the carrier occupation
far from the semiconductor-insulator interface, where $V(x)=0$. 
The variations of  $V(x)$ and $\delta(x)$ with 
the distance $x$ are determined by the Poisson equation. 
For the accumulation layer, where $\delta(x) \gg \delta_0$, 
the following relation between the electric field
$F(x) = - dV(x)/dx$ and $\delta(x)$ can be obtained,~\cite{horowitz} 
\begin{equation}
F^2(x)= 2 k_{\rm B} T_0 N_{\rm t} \delta(x) / \epsilon_{\rm s}, 
\end{equation}
where $\epsilon_{\rm s}$ is the dielectric constant of the semiconductor. 
The field $F(0)$ at the interface can be expressed in terms of the 
gate voltage $V_{\rm G}$ and and the insulator capacitance per 
unit area $C_{\rm i}$ through Gauss' law, 
\begin{equation}
F(0) = C_{\rm i} V_{\rm G} / \epsilon_{\rm s}.
\end{equation}

Substituting the distance-dependent charge occupation $\delta(x)$ into 
Eq.~(\ref{eq:sigmaqT}) for the conductivity,  
the source-drain current of the transistor in the linear regime
($-V_{\rm D} < -V_{\rm G}$) reads
\begin{equation}
I =  \frac{W V_{\rm D}}{L} \int^t_0 dx\: 
\sigma \left[ \delta \left( x \right), T \right] .
\end{equation}
Here, $V_{\rm D}$ is the drain voltage (the source is the ground electrode) 
and $L$, $W$ and $t$ are the length, 
width, and thickness of the channel, respectively. 
The field-effect mobility then follows from the 
transconductance (see, e.g., Ref.~\onlinecite{overview})
\begin{equation}
\mu_{\rm FE} \equiv \frac{L}{C_{\rm i} W V_{\rm D}} 
\:\frac{\partial I}{\partial V_{\rm G}}.
\label{mu_lin}
\end{equation}
From Eqs.~(\ref{eq:sigmaqT})--(\ref{mu_lin}) 
we obtain the following expression 
for the field-effect mobility, 
\begin{equation}
\mu_{\rm FE} = \frac{\sigma_0}{e}  
\left(\frac{\pi  \left(T_0/T\right)^3}
{\left(2 \alpha\right)^3  B_{\rm c} \:\Gamma(1-T/T_0) \:\Gamma(1+T/T_0)}
\right)^{T_0/T} 
{\left[ \frac{{\left( C_{\rm i} V_{\rm G} \right)}^2}{ 
2 k_{\rm B} T_0 \: \epsilon_{\rm s}} \right]}^{T_0/T - 1}, 
\label{mobility}
\end{equation}
where we have assumed that the thickness $t$ of the
semiconductor layer is sufficiently large such that $V(t)=0$. 
Then the field-effect mobility is independent of the thickness
 $t$ as well as the bulk carrier occupation $q_0$.
We note that the $N_{\rm t}$ dependence of the charge distribution 
in the accumulation layer is exactly cancelled by the 
$N_{\rm t}$ dependence of $\sigma(q,T)$. 

\begin{figure}
\begin{center}
\mbox{\psfig{file=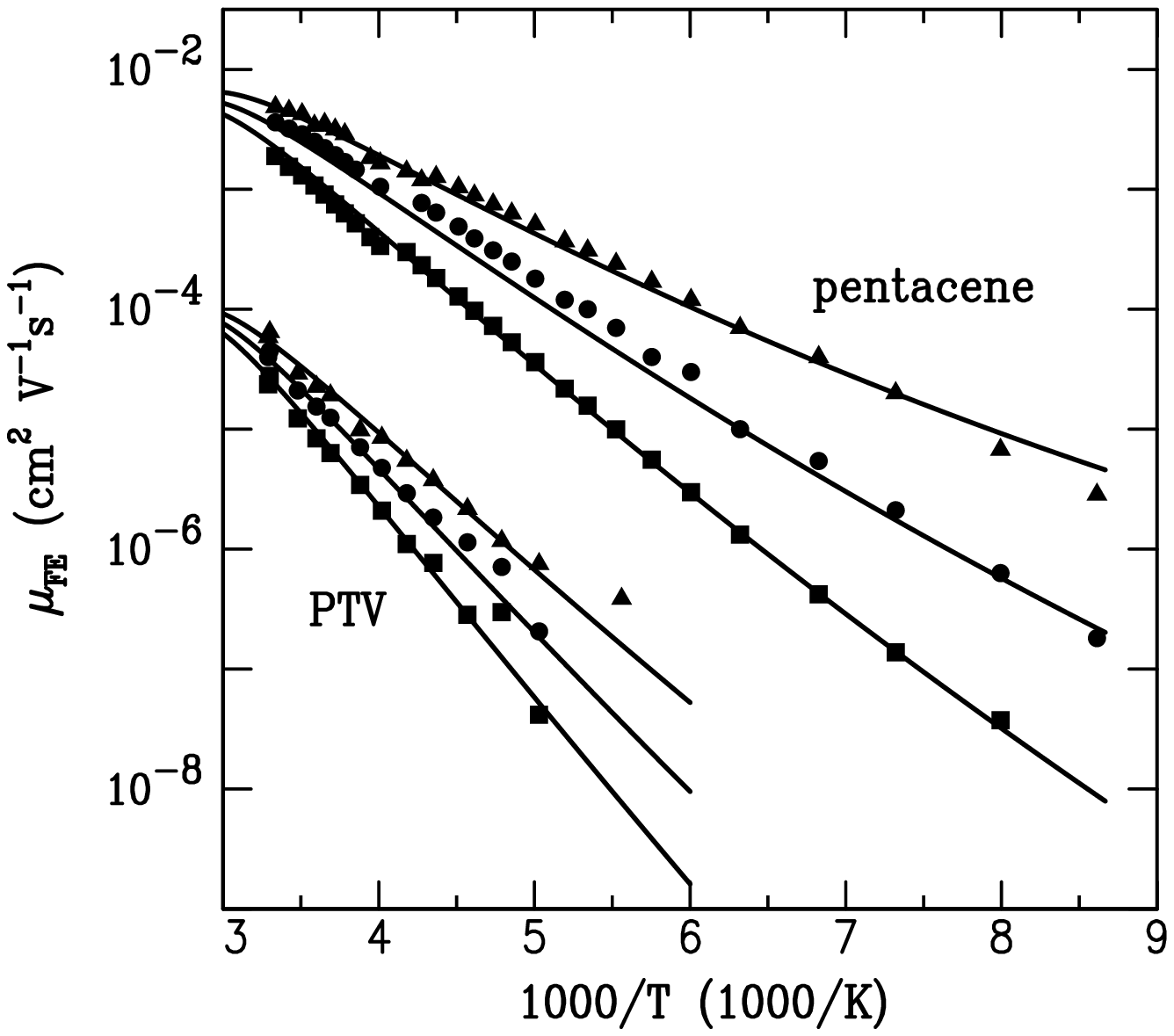,angle=0,width=7.4cm}}
\end{center}
\caption[]{Field-effect mobility $\mu_{\rm FE}$
 in a pentacene and a polythienylene vinylene (PTV)
thin-film transistor as a function of the temperature $T$
 for different gate voltages
$V_{\rm G}$ = -20 $V$ (triangles), -10 $V$ (circles) and -5 $V$ (squares). 
The experimental data (symbols) are taken from
Ref.~\onlinecite{overview}. 
The solid lines are according to Eq.~(\ref{mobility}), 
using the parameters given in Table~\ref{tabel}.}  
\label{data} 
\end{figure}

\begin{figure}
\begin{center}
\mbox{\psfig{file=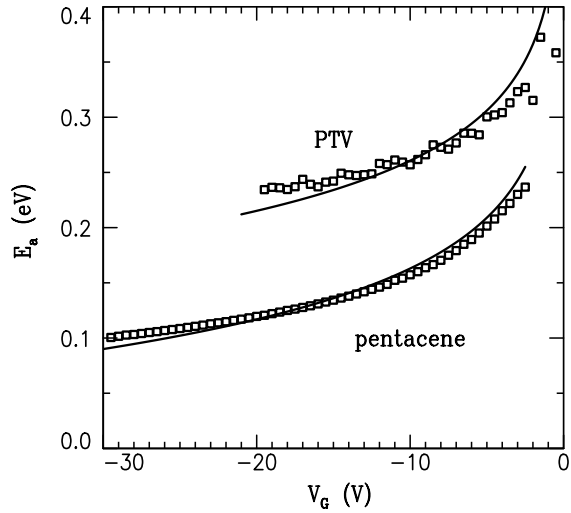,angle=0,width=7.4cm}}
\end{center}
\caption[]{Activation energy $E_{\rm a}$ for the field-effect mobility 
 in a pentacene and a polythienylene vinylene (PTV)
thin-film transistor as a function of the 
 gate voltage $V_{\rm G}$.
The experimental data (squares) are taken from Ref.~\onlinecite{overview}.
The solid lines are calculated from Eq.~(\ref{mobility}), 
using the parameters given in Table~\ref{tabel}.}
\label{activering}
\end{figure}

Let us now apply our result~(\ref{mobility}) 
to the experimental data of Ref.~\onlinecite{overview}, where
 the drain current $I$ versus gate voltage $V_{\rm G}$ 
characteristics have been measured of 
both a pentacene and a polythienylene vinylene (PTV) 
organic thin-film transistor at a range of temperatures. 
The precursors of both organic semiconductors 
are spin-coated from solution on a substrate consisting of a 
heavily {\it n}-doped silicon (common) gate electrode, 
a 200 nm thick SiO$_2$ insulating layer 
($C_{\rm i}=17$ nFcm$^{-2}$) 
and a patterned gold layer 
as the source and drain electrodes. 
The precursors are converted into the organic
semiconductors using a process described in Ref.~\onlinecite{overview}. 
Typical channel widths and lengths were $W=10$-20 mm and 
$L=2$-20 $\mu$m respectively. 
The film thickness $t$ varied from 30 to 50 nm. 
For both semiconductors, we use a relative dielectric constant
$\epsilon_{\rm r}=3$, which is appropriate for most organic solids.
In Fig.~\ref{data} the field-effect mobility in
a pentacene and in a PTV thin-film transistor is plotted 
 against the inverse temperature for different gate
voltages. 
Experimentally, the field-effect mobilities are determined from  
Eq.~(\ref{mu_lin}) at $V_{\rm D}=-2$ V.
The theoretical curves (solid lines) follow from Eq.~(\ref{mobility}), 
where we have used $\sigma_0$, $\alpha$, and $T_0$ as fitting 
parameters.  
The agreement with experiment is quite good  
 (the parameter values are given in Table~\ref{tabel}). 
The temperature dependence of $\mu_{\rm FE}$, as shown in Fig.~\ref{data}, 
follows a simple Arrhenius behavior 
$\mu_{\rm FE} \sim \exp{\left[ - E_{\rm a} / ( k_{\rm B} T ) \right]}$, where
the activation energy $E_{\rm a}$ depends on $V_{\rm G}$  
as plotted in Fig.~\ref{activering}. 
The decrease of  $E_{\rm a}$ with increasing (negative) 
gate voltage is the direct result of accumulated charges filling the 
lower-lying states. 
As a result, any additional charge carriers in the system will 
occupy sites with---on average---a higher energy and less energy 
will be required for the activated jumps to neighboring sites. 

The field-effect mobility in PTV is more than two orders of magnitude 
lower than the field-effect mobility in pentacene. 
Furthermore, the activation energy for PTV is about twice 
the activation energy for pentacene. 
Surprisingly, these differences cannot be attributed to 
differences in the prefactor $\sigma_0$ nor to the 
width of the energy distribution $T_0$, as these parameters 
have similar values for PTV as well as pentacene (see Table~\ref{tabel}). 
The main difference between
pentacene and PTV appears to be in the overlap parameter $\alpha$, 
which determines the tunneling process between different sites. 
We note that this key parameter is absent in 
a multiple-trapping model, where the transport is 
governed by thermal activation from traps to a conduction band 
and subsequent retrapping, without involving a tunneling step. 
As the length scale $\alpha^{-1}$ is smaller than the size of a molecule,   
 one must be cautious not to interpret 
$\alpha^{-1}$ simply as the decay length of the electronic wave function. 
The size and shape of the molecules and the morphology of the 
organic film are expected to have an important influence 
on the tunneling probability as well. 
The observed difference in 
$\alpha^{-1}$ may be due to the fact that there is more 
steric hindrance in the polymer PTV than in a system of small 
pentacene molecules. The better stacking properties of 
pentacene give rise to a larger area of overlap of the electronic 
wave functions, which results in a larger effective overlap  
$\alpha^{-1}$ in our model. 

In conclusion, we have derived an 
analytic expression for the field-effect mobility in a
thin-film transistor of an amorphous organic semiconductor, 
using percolation theory and the concept of hopping in an
exponential density of localized states. 
The calculated temperature dependence and gate-voltage dependence 
agree well with those of the  
the observed field-effect mobility in both a 
pentacene and a PTV thin-film transistor. 
According to the theory, the differences in the magnitude 
and in the temperature dependence of the field-effect mobility 
of pentacene and  PTV transistors are 
mainly due to differences in the
structural order of the organic films.

We would like to thank A. R. Brown, C. P. Jarrett, and D. M. de Leeuw
 for providing the data of Fig~\ref{data}, and 
M. J. M. de Jong for useful discussions. 
Financial support from the EC under Esprit project 24793 Frequent 
and from the Dutch Science Foundation NWO/FOM is gratefully 
acknowledged.

\begin{table}
\caption{The pre-exponential factor to the conductivity $\sigma_0$, 
the overlap parameter $\alpha^{-1}$, and the width 
of the exponential distribution of localized states $T_0$
for both pentacene and polythienylene vinylene (PTV)
as obtained from the fit of  Eq.~(\ref{mobility})
to the experimental data of Ref.~\protect\onlinecite{overview}, 
see Fig.~\ref{data}.}
\[
\begin{array}{cccc}
\hline\hline
& \; \sigma_0 \; \; (10^{10} \; {\rm S/m)} \; 
& \; \alpha^{-1} \; ({\rm \AA}) \;
& \; T_0 \; ({\rm K}) \; \\
\hline
\; {\rm pentacene} \; & 1.6            & 2.2       & 385 \\
\hline 
 \; {\rm PTV}   \;    & 0.7            & 0.8       & 380 \\
\hline \hline
\end{array}
\]
\label{tabel}

\end{table}

\end{document}